\documentclass{egpubl}
\usepackage{eg2026}

\ConferencePaper        
\CGFStandardLicense

\usepackage[T1]{fontenc}
\usepackage{dfadobe}

\usepackage{cite}  
\BibtexOrBiblatex
\electronicVersion
\PrintedOrElectronic
\ifpdf \usepackage[pdftex]{graphicx} \pdfcompresslevel=9
\else \usepackage[dvips]{graphicx} \fi

\usepackage{egweblnk}

\usepackage{bm}

\usepackage{cancel}

\usepackage[textsize=tiny]{todonotes}

\usepackage[labelformat=simple]{subcaption}
\usepackage{amsmath}
\usepackage{float}
\usepackage{amssymb}
\usepackage{soulutf8}
\sethlcolor{blue!20}

\usepackage{mathrsfs}

\usepackage{multirow} 
\usepackage{makecell} 
\usepackage{utfsym} 
\usepackage{hyphenat} 

\usepackage{booktabs} 
\usepackage[normalem]{ulem}
\usepackage{enumitem}

\usepackage[ruled]{algorithm2e} 

\SetAlFnt{\small}
\SetAlCapFnt{\small}
\SetAlCapNameFnt{\small}
\SetAlCapHSkip{0pt}

\usepackage{cases}

\usepackage{xspace}

\newcommand{\keep}[1]{}
\newcommand{\old}[1]{}

\newcommand{\fig}{Figure{}~}

\title[DexterCap]%
      {DexterCap: Affordable and Automated Capture of Complex Hand-Object Interactions}

\author[Y. Liang et al.]{
\parbox{\textwidth}{\centering
    Yutong Liang\thanks{These authors contributed equally to this work.}$^{1}$\orcid{0009-0001-8448-4510}
    and Shiyi Xu\footnotemark[1]$^{1}$\orcid{0009-0004-9697-6565}
    and Yulong Zhang\footnotemark[1]$^{2}$\orcid{0000-0002-2353-1416}
    and Bowen Zhan$^{1}$\orcid{0009-0008-4350-5427}
    and He Zhang$^{3}$\orcid{0009-0003-8607-9441}
    and Libin Liu\thanks{Corresponding author.}$^{1}$\orcid{0000-0003-2280-6817}
}
\\
{
\parbox{\textwidth}{\centering
    $^1$Peking University, Beijing, China\\
    $^2$School of Computer Science, Peking University, Beijing, China\\
    $^3$Tencent Robotics X, Shenzhen, China}
}
}

\begin{document}

\teaser{
    \label{fig:teaser}
    \centering
    \includegraphics[width=0.9\linewidth]{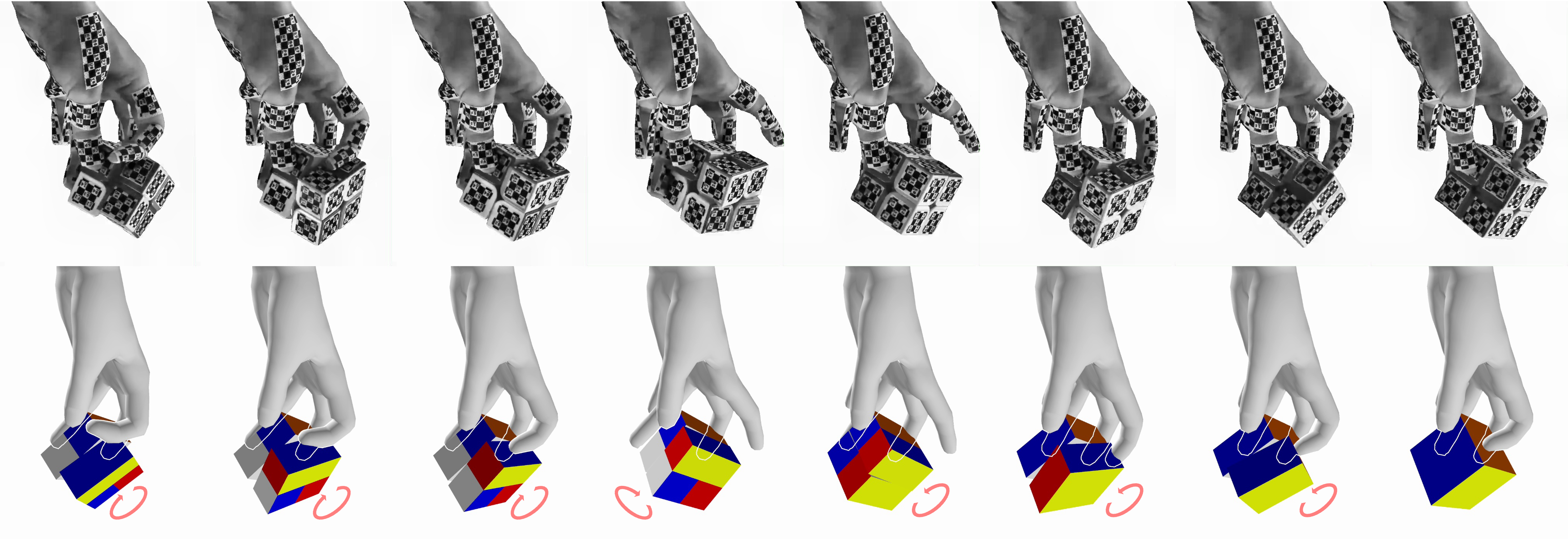}
    \caption{DexterCap captures dexterous manipulation of a Rubik’s Cube. Top: raw multi-camera footage showing character-coded marker patches. Bottom: reconstructed 3D hand and object motion. Dense marker coverage enables accurate reconstruction of subtle, natural in-hand manipulation motions.}
}

\maketitle
\begin{abstract}

Capturing fine-grained hand-object interactions is challenging due to severe self-occlusion from closely spaced fingers and the subtlety of in-hand manipulation motions. Existing optical motion capture systems rely on expensive camera setups and extensive manual post-processing, while low-cost vision-based methods often suffer from reduced accuracy and reliability under occlusion. To address these challenges, we present DexterCap, a low-cost optical capture system for dexterous in-hand manipulation. DexterCap uses dense, character-coded marker patches to achieve robust tracking under severe self-occlusion, together with an automated reconstruction pipeline that requires minimal manual effort. With DexterCap, we introduce DexterHand, a dataset of fine-grained hand-object interactions covering diverse manipulation behaviors and objects, from simple primitives to complex articulated objects such as a Rubik's Cube. We release the dataset and code to support future research on dexterous hand-object interaction. Project website: \url{https://pku-mocca.github.io/Dextercap-Page/}


\end{abstract}

\section{Introduction}

\label{sec:introduction}
Human hands, with their flexible structure and high degrees of freedom, play a central role in interacting with the physical world. They perform a wide range of dexterous actions that support everything from routine daily tasks to complex, specialized activities, using skills that range from simple grasping and moving of objects to in-hand manipulations involved in writing, assembling components, or handling tools in precision manufacturing.

Despite its importance, the precise capture and realistic modeling of fine-grained hand skills remain a significant challenge. Although human-object interaction (HOI) datasets and models are becoming increasingly available, most existing work focuses on basic actions such as grasping and moving, or interactions with simple articulated objects~\cite{taheri2020grab,fan2023arctic,kim2024parahome}. There is a substantial gap in resources for understanding and synthesizing the nuanced dexterity required for a wide range of hand skills.

Current motion capture technologies, whether marker-based optical systems, data gloves, or marker-less vision-based approaches, each present certain limitations. Marker-based optical systems, although accurate, typically rely on expensive camera setups and physical markers. A commonly used setup involves homogeneous markers, which can lead to issues such as shifting and mislabeling. These problems often require extensive manual identification and cleaning during post-processing. These challenges become more pronounced when capturing hand manipulation, where markers are placed in close proximity and occlusion frequently occurs. 
Data gloves typically rely on inertial sensors and are often affected by inherent error accumulation, primarily caused by gyroscope drift and accelerometer bias, which results in progressive deviations in finger and hand pose estimation over time. Marker-less vision-based approaches offer greater flexibility, affordability, and scalability. However, in the absence of explicit visual markers, they often struggle to achieve the accuracy and robustness needed to capture subtle finger movements during object manipulation.

To address these challenges, this paper introduces \emph{DexterCap}, a comprehensive framework for capturing, solving, and reconstructing hand and object motions in dexterous hand-object interactions. Our goal is to build an affordable system that enables fast acquisition of a large-scale and diverse dataset of HOI skills, while ensuring accurate recovery of fine-grained finger motions. The affordability of the system is twofold. First, it is built using low-cost hardware, making the setup economically accessible. Second, after a one-time preparation for camera calibration and MANO shape estimation, the pipeline runs largely automatically. While a small number of frames must be labeled initially to train the image-processing models, as discussed in Sec.~\ref{ssec:system_setup}, these models generalize well to new sessions with data accumulated from previous captures, making subsequent solving fully automatic, substantially reducing human effort and enabling scalable data collection.

Our system draws inspiration from the vision-based approach introduced by~\cite{chen2021capturing} for capturing human body deformation during full-body motion. They demonstrated that deep learning-based computer vision techniques, combined with specially designed visual patterns, enables the automated identification and localization of a dense set of markers. Building on this idea for dexterous hand-object manipulation, we present a practical, low-cost methodology that operates with affordable hardware and minimal labeling, applies to arbitrary rigid objects, and is readily extendable to articulated ones. By leveraging a large number of tagged markers, a system can enhance robustness against occlusion and significantly reduce the need for tedious manual cleanup.

However, to design such a motion capture system for hands, we need to address two challenges: (1) limited resolution requires placing the camera close to the hand, causing limited depth of field and variation in hand scale within the image as it moves, which complicates robust marker detection across multiple scales; (2) markers may be missing in varying patterns due to occlusion and noise, making it difficult to recover the full poses of the hand, fingers, and object, while also requiring careful calibration.

We address the first challenge by introducing a three-stage marker detection and identification model that cascades the detection of markers, edges, and tags, enabling robust and efficient marker recognition. To tackle the second challenge, we employ the morphable hand model MANO~\cite{romero2017embodied} and develop a robust calibration and solving procedure to ensure accurate reconstruction of both hand and object poses.

To demonstrate the performance of \emph{DexterCap}, we introduce \emph{DexterHand}, an open-source HOI dataset specifically designed to capture complex in-hand object manipulation. It includes a diverse set of manipulation skills performed on both primitive shapes and articulated objects, such as a Rubik's Cube. The motions are captured with high detail, precise control, and extended duration, with most sequences lasting over 10 minutes. We evaluate both \emph{DexterCap} and \emph{DexterHand} using a set of quantitative metrics and benchmarks. All tools, models, and data, including the full dataset and our processing codebase, will be released to support future research.

\section{Related Work}

\label{sec:related_work}

\subsection{Motion Capture for Human-Object Interaction}

Accurately capturing human motion, particularly of the hands, is fundamental to understanding and generating dexterous manipulation and realistic animations. Various techniques have evolved, with optical, inertial, and markerless vision-based systems being the most prominent.

Marker-based optical motion capture is often considered the high-precision ``gold standard'', with systems like Qualisys and Vicon widely used. These systems use multiple cameras to track markers, enabling sub-millimeter-accurate 3D reconstruction from high-frame-rate positional data~\cite{lu2025humoto,kim2024parahome,han2018online}.
Despite their accuracy, these systems are costly and prone to occlusion issues (e.g., marker swapping), which often necessitates extensive manual post-processing to ensure the quality of captured motion data.
Novel solutions include T-pose initialization~\cite{meyer2014online} and learning-based correspondence methods that use synthetic data and neural networks to improve marker solving robustness~\cite{holden2018robust,han2018online,chen2021mocap,ghorbani2021soma,pan2023locality}. Other approaches leverage monocular video cues~\cite{milef2024towards} or employ unique markers~\cite{chen2021capturing}.
However, when extending such systems to more interaction scenarios, particularly those involving multi-jointed objects like a Rubik’s cube, a tedious manual process is still required.

Sensor-based methods, using  Inertial Measurement Units (IMUs) or joint force contact sensors, are less prone to visual occlusion and more cost-effective.  To further improve the usability,  recent research has explored reducing the number of IMUs in TransPose~\cite{yi2021transpose} and PIP~\cite{yi2022physical}, and enhancing accuracy by mitigating issues such as local inertial forces~\cite{yi2024physical}. 
However, capturing fine-grained hand motions with IMU gloves still remains challenging in terms of finger accuracy. Moreover, in interaction-rich scenarios, the introduction of objects typically requires additional markers and alignment ~\cite{kim2024parahome}.

Vision-based markerless methods estimate body or hand pose directly from images or videos, eliminating the need for markers or specialized suits and thus offering greater convenience and lower cost~\cite{elmoubtahij2022structured}. These approaches leverage various imaging modalities, including single/multiple RGB cameras~\cite{pavlakos2024reconstructing, potamias2024wilor}, depth cameras~\cite{hu2022physical}, infrared cameras~\cite{baker2025optimisation}, with multi-view systems alleviating occlusion and enhancing 3D accuracy.
Despite these advances, vision-based markerless methods still face persistent challenges, including severe occlusions, lighting variations, motion blur, and maintaining temporal consistency~\cite{elmoubtahij2022structured}. 
Furthermore, although there are recent works that incorporate reinforcement learning to enforce realistic hand-object interactions~\cite{haoyu2024hoic}, ensuring the physical plausibility of the estimated poses still remains difficult.

\subsection{Hand-Object Interaction Datasets}

Existing human-object interaction datasets primarily focus on capturing coordinated whole-body interactions with objects, but often lack precise details regarding hand-object contact ~\cite{MOGAZE, BEHAVE, OMOMO, SEMANTIC-HOI, HODome, CIRCLE}. 
The integration of optical motion capture systems has enabled the collection of more detailed grasping motions ~\cite{taheri2020grab, FLEX, fu2025gigahands}.  Recent work expands these datasets to include a wider variety of bimanual functional manipulations ~\cite{HIMO} and interactions with articulated objects ~\cite{fan2023arctic, OAKINK2, CHOICE}. Instead of focusing on whole-body interaction with broad hand motion, our dataset specifically targets fine-grained in-hand manipulation.

To capture hand-object interactions, many datasets rely on RGB-D reconstruction, which, while generalizable, often falls short in recording detailed hand movements ~\cite{OBMAN, hampali2020honnotate, DexYCB, H2O, HOI4D, OakInk}. Datasets such as ContactPose ~\cite{CONTACTPOSE} emphasize the collection of affordance ~\cite{YCB-AFFORDANCE} and contact information, which can improve grasp accuracy but still lack dynamic hand motion details. While ManipNet ~\cite{MANIPNET} includes precise in-hand manipulations, such as twisting and rotating ~\cite{INTERHAND26M}, its object coverage remains limited to rigid items. Thanks to our system design and specialized marker patches, our dataset captures accurate in-hand contact and subtle hand motions. Additionally, we include complex object interactions, such as those involving a Rubik’s cube, in our dataset.

\section{Dexterous Hand Motion Capture System}
\label{sec:system}

Our system, \emph{DexterCap}, is a multi-camera, vision-based motion capture and reconstruction framework designed to achieve high-precision tracking of hand and object motions in dexterous in-hand interactions. It offers a relatively low-cost solution with an easily deployable and automated hardware-software setup. The system consists of a set of synchronized industrial cameras positioned around the capture space from multiple directions. We attach a set of specially designed visual markers to the subject's hand and the object. Hand motions are captured simultaneously by all cameras. These visual markers are then detected in the recorded videos with minimal manual labeling, aided by learning-based marker detection models. Finally, hand and finger motions are estimated using the morphable hand model MANO~\cite{romero2017embodied}, while object poses are recovered using object-specific solvers. In this section, we detail the system design and associated algorithms.

\subsection{Hardware Setup and Marker Design}
\label{ssec:hardware_setup}

\subsubsection{Cameras}
Our current \emph{DexterCap} setup uses a set of synchronized industrial cameras that record grayscale video at a resolution of $2048\times{}2448$ and 20~FPS. The cameras are mounted on a $2\times{}1\times{}2$~m cage and are arranged to minimize occlusions and ensure robust 3D hand tracking. All cameras are calibrated using the standard procedure with the OpenCV library~\cite{opencv_library}. A short exposure time of 1ms is used to prevent motion blur, and the aperture is narrowed as much as possible to increase depth of field. The system is installed in a bright room with natural lighting, supplemented by a small set of LED lights that provide additional illumination when needed.

\subsubsection{Markers}

\begin{figure}[t]
    \centering
    \begin{subfigure}{0.49\linewidth}
        \includegraphics[width=1.0\linewidth]{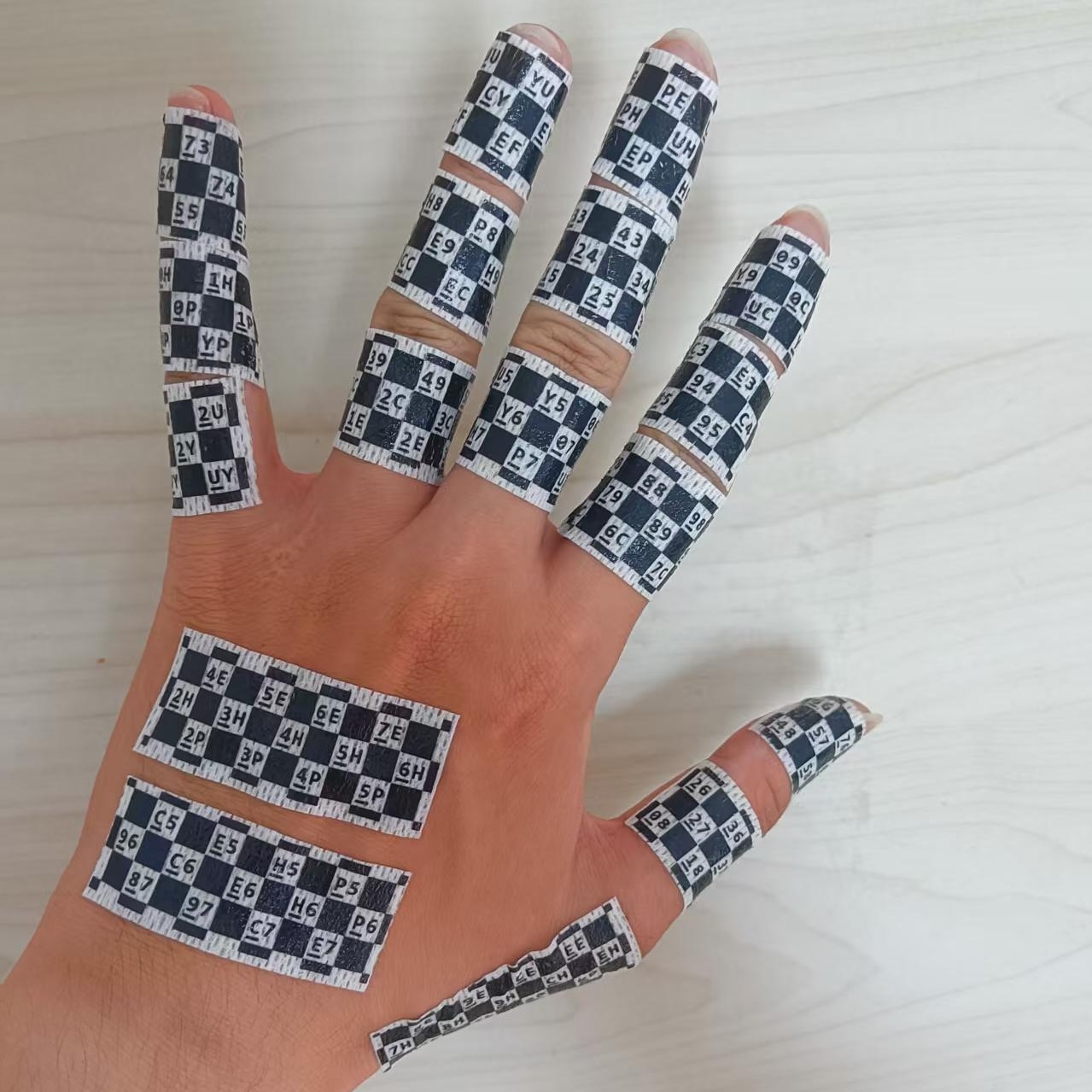}
        \caption{Markers affixed to the left hand.}
    \end{subfigure}
    \begin{subfigure}{0.49\linewidth}
        \includegraphics[width=1.0\linewidth]{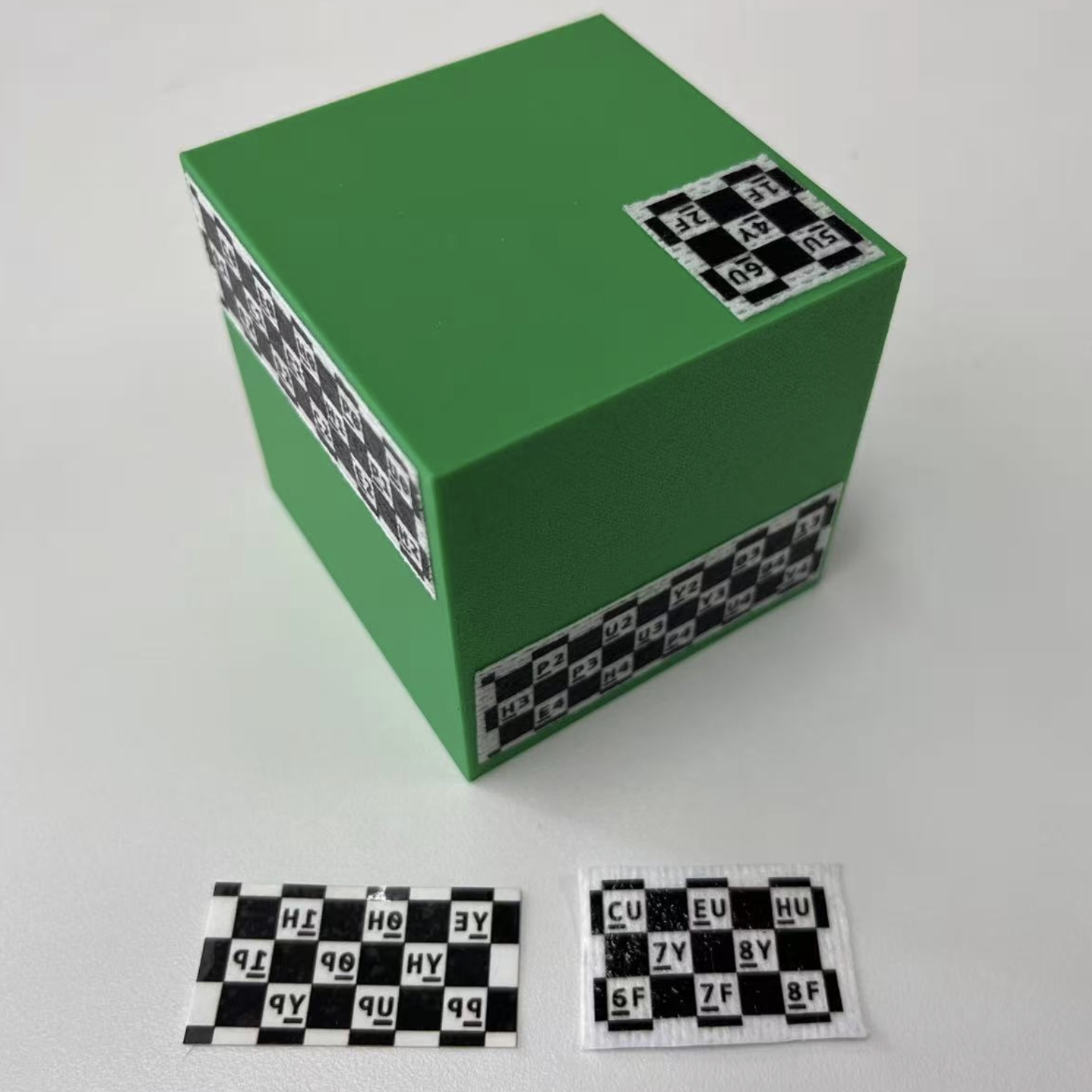}
        \caption{Object marker application.}
    \end{subfigure}
    \caption{
        Our marker system: 
        (a) Markers are attached to rigid hand regions (phalanges, dorsum) for accurate tracking. 
        (b) Top: Rigid object (green cube) with visual marker patches. 
        Bottom left: A visual marker patch printed on a transfer sticker.
        Bottom right: Visual markers transferred onto a medical tape for secure attachment. 
    }
    \label{fig:markers_system}
\end{figure}

Inspired by~\cite{chen2021capturing}, we use a customized marker system with high-contrast checkerboard patterns for robust tracking and identification. As illustrated in \fig\ref{fig:markers_system}, each white square contains a unique two-character ID in black, selected from 26 uppercase letters and 10 digits, excluding visually similar characters, yielding 324 unique tags. We place an underscore below the left character to help determine orientation.

\begin{figure*}[t]
    \centering
    \begin{subfigure}{0.24\linewidth}
    \includegraphics[width=\linewidth,height=4cm,keepaspectratio]{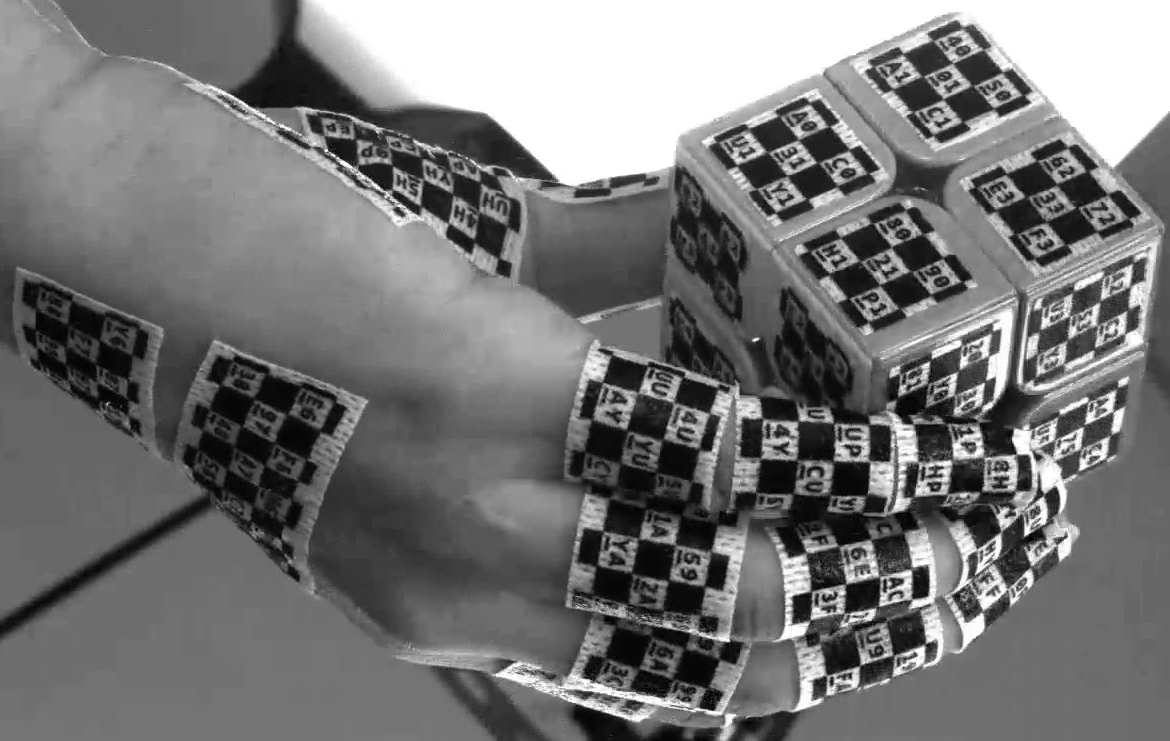}
    \caption{Raw input image.}
    \end{subfigure}
    \begin{subfigure}{0.24\linewidth}
    \includegraphics[width=\linewidth,height=4cm,keepaspectratio]{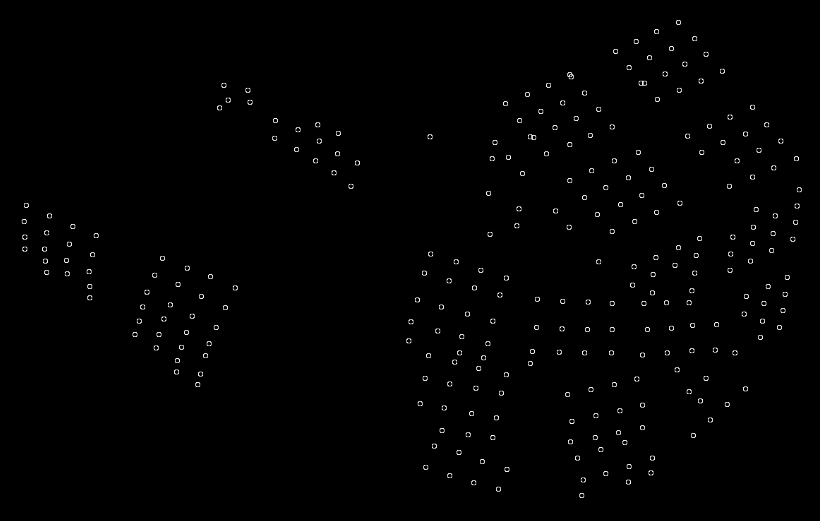}
    \caption{CornerNet detection.}
    \end{subfigure}
    \begin{subfigure}{0.241\linewidth}
    \centering
    \includegraphics[width=\linewidth,height=4cm,keepaspectratio]{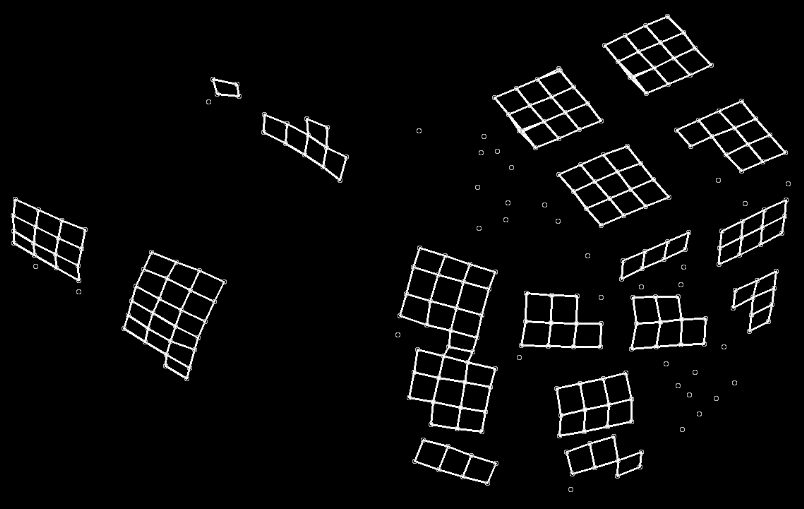}
    \caption{EdgeNet classification.}
    \end{subfigure}
    \begin{subfigure}{0.251\linewidth}
    \includegraphics[width=\linewidth,height=4cm,keepaspectratio]{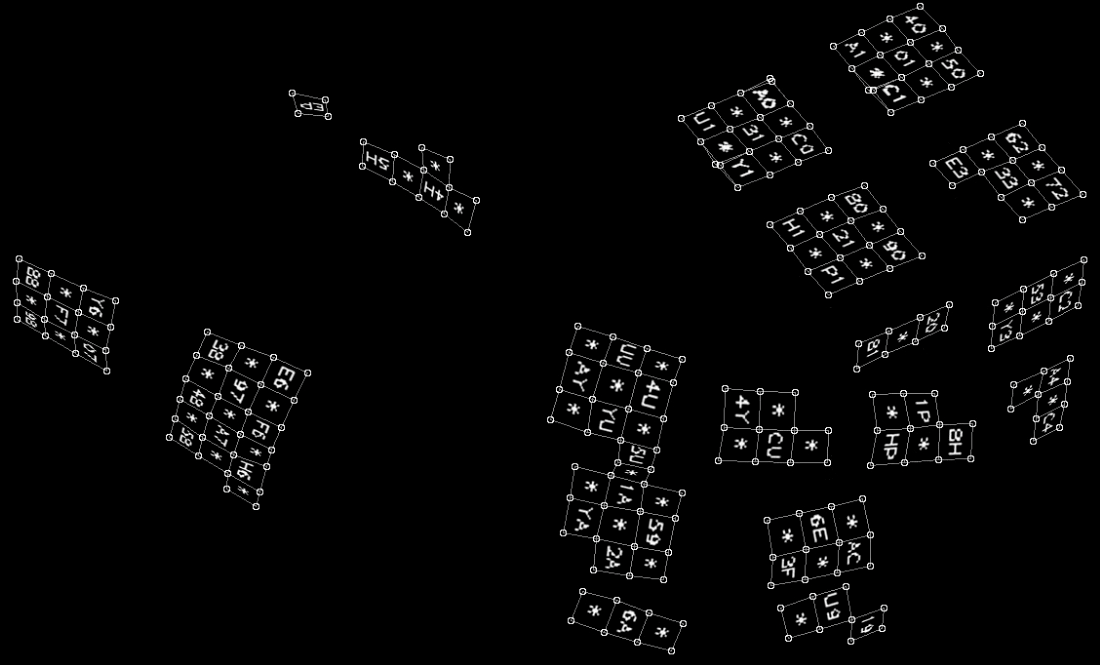}
    \caption{BlockNet recognition.}
    \end{subfigure}
    \caption{Image processing pipeline for marker detection.\ (a) Raw input image with character-coded checkerboard markers.\ (b) CornerNet detection results.\ (c) EdgeNet edge classification results.\ (d) BlockNet block recognition with character identifiers.}
    \label{fig:marker_detection_pipeline}
\end{figure*}

\cite{chen2021capturing} created a special suit with visual patterns printed directly onto the fabric. However, we found that the equivalent idea using a glove does not work well for hand motion capture. A glove can stretch and wrinkle under different hand poses and may slide along the skin, which introduces errors in pose estimation. Instead, we attach marker patches directly to each relatively rigid region of the hand, including the finger knuckles, the dorsum, and the palm, for a total of 19 patches per hand. These patches are transfer printed onto medical adhesive tape for reliable and comfortable attachment. This dense marker layout provides over 500 detectable corners, ensuring accurate motion reconstruction even under severe occlusion. Markers are also placed on object surfaces to support robust pose estimation for both rigid and articulated objects such as a Rubik's Cube.

\subsection{Image Processing and Marker Extraction}
\label{sec:image_processing}

Once the raw video data is acquired, each frame is processed through our image processing pipeline to detect and identify visual markers. As illustrated in \fig\ref{fig:marker_detection_pipeline}, the pipeline first detects candidate corners, then identifies edges between them to assemble blocks, and finally recognizes the tag of each block, which encode the corner IDs. Commonly used marker systems, such as ArUco~\cite{ArUco} from OpenCV, often perform poorly in this setting due to severe deformation of the markers. To address this, we train dedicated models for marker detection.

\subsubsection{Corner Detection}
\label{sssec:corner_detection}

Precise localization of the corners in the checkerboard-style marker pattern is critical to our system. To achieve this, we adopt a deep learning approach based on the U-Net~\cite{ronneberger2015unet} architecture, which we refer to as \emph{CornerNet} in our work. CornerNet takes a $64 \times 64$ grayscale image as input and produces a heatmap of the same resolution, where each corner is represented by a Gaussian peak. The raw input images are divided into overlapping patches of $64 \times 64$ pixel for efficient processing. For details of CornerNet, please refer to Appendix\mbox{~\ref{sec:network_architecture}}. To handle scale variation, we also downsample the images by half and process them in the same manner, which provides an expanded field of view and improves robustness to perspective changes. Compared with direct coordinate regression~\cite{chen2021capturing}, this heatmap-based formulation is more reliable at both patch and image levels (see Sec.~\ref{ssec:recognition_performance}). It also allows each patch to contain multiple corners and avoids ambiguities when merging overlapping predictions.

We train CornerNet by minimizing the Mean Squared Error between the predicted heatmap and the ground truth. During inference, the full image is divided into patches and processed in batches. The resulting heatmap patches are stitched together, and overlapping regions are merged using a maximum value strategy. After normalization, we apply a relatively low confidence threshold of 0.6 to binarize the heatmap, which is chosen to favor high recall and avoid missing true corners at this stage. Potential false positives are subsequently pruned by the edge detection stages. Finally, corners are identified as the centers of the connected components in the binary map.

\subsubsection{Edge and Block Detection}
\label{sssec:edge_block_recognition}

With the candidate corners detected, the next step is to determine how they assemble into checkerboard blocks. \cite{chen2021capturing} proposed validating all candidate quads generated based on geometric criteria. However, we find these criteria difficult to configure, and they often produce a large number of candidate quads, resulting in high computational cost during block validation. To address this, we adopt a two-stage block detection strategy. First, we check whether each pair of candidate corners is directly connected in the template pattern, and then assemble blocks based on these connections.

Specifically, we employ a neural network, \emph{EdgeNet}, based on ResNet~\cite{he2015resnet} with a fully connected classification head that acts as a binary classifier to determine whether a valid edge exists between a pair of candidate corners. For each pair of candidate corners within a predefined Euclidean distance threshold, we extract a rectangular image patch centered on the segment connecting the two corners and aligned with its direction. The distance threshold is determined from the statistics of edge lengths in the manually annotated training set and is chosen to achieve 95\% recall. This choice balances accuracy and computational efficiency by excluding implausibly long connections while preserving the majority of valid edges. The extracted patch is resized to $64 \times 64$ and used as input to EdgeNet. Once all connections are validated, we perform a graph search to identify corners and edges that form convex quadrilateral blocks with acceptable aspect ratios, avoiding shapes that are excessively thin or distorted.

EdgeNet is trained using a Binary Cross-Entropy (BCE) loss. To construct the training set, we employ a balanced sampling strategy that includes correct edges (40\%), block diagonals (15\%), lines between random corners (30\%), and random point pairs (15\%). During inference, we apply a conservative confidence threshold of 0.75 to prioritize precision, thereby preventing incorrect block assembly while retaining sufficient connectivity. Candidate corners that do not belong to any detected block are discarded. At this stage, we do not require the detected blocks to be perfectly valid, as they will be further validated in the subsequent stage.

This edge-first assembly prunes obviously invalid candidates early and avoids redundant computation on shared edges, shrinking candidates by an order of magnitude and improving precision/recall over \cite{chen2021capturing} (see Sec.~\ref{ssec:recognition_performance}).

\subsubsection{Block Identification}
The final stage of the image processing pipeline is to validate and identify the detected blocks from the previous steps. Following \cite{chen2021capturing}, we employ a neural network, \emph{BlockNet}, for this task. Our BlockNet is a ResNet-based architecture with three output heads, each responsible for predicting one character of the tag and the orientation of the block. We use one-hot encoding for both the tag letters and the orientation, which allows BlockNet to be effectively trained using the standard cross-entropy loss. At inference time, blocks that are determined to be invalid, along with their corresponding corners, are discarded.

\subsubsection{Post-Processing}
To enhance robustness, we employ a post-processing procedure based on a voting mechanism that leverages the marker pattern defined in the template to correct potential mislabeling. Each candidate block votes on the tags of all other blocks within the same marker patch. For each block, the tag with the highest vote count is selected as its final prediction. Once the tags of the blocks are identified, each detected corner can be assigned a unique ID, which is used to match corresponding corners across different images.

\subsection{3D Marker Reconstruction}
\label{ssec:3d_reconstruction}

After extracting 2D marker coordinates and IDs from multiple views, we reconstruct their 3D positions using triangulation. Only markers observed by at least three cameras are considered. RANSAC is also applied during triangulation to reject outlier observations.

The initial 3D points may contain outliers due to incorrect marker identification. We detect and remove these outliers using two heuristics. First, for each patch in the template marker patterns (as shown in Figure~\ref{fig:markers_system}), we cluster the points within the patch based on their 3D distances. The cluster with the largest number of points is retained, while all other points are treated as outliers and discarded. Second, we compute the z-scores of each point's coordinates within a sliding window and remove points whose scores exceed a predefined threshold.

After removing outliers, we fill missing markers using a simple strategy: the position of a missing marker at frame at frame $i$ is estimated using linear interpolation when the marker is observed in at least one of the two preceding frames ($i-1$ or $i-2$) and at least one of the two subsequent frames ($i+1$ or $i+2$).

\subsection{Hand Motion Reconstruction}

With the 3D markers reconstructed from the previous step, we proceed to solve for hand and object motions. We use the parametric MANO model~\cite{romero2017embodied} as the hand template. In this section, we describe how we adapt the model, perform calibration, and solve for hand motion.

\subsubsection{Hand Model Parameterization}
\label{sec:hand_model_construction}
The MANO model represents a 3D hand mesh using two sets of parameters $(\beta, \theta)$: $\beta \in \mathbb{R}^{10}$, which controls the shape of the hand, and $\theta \in \mathbb{R}^{45}$, which defines the pose through joint rotations. 

\begin{figure}[t]
    \centering
    \includegraphics[width=0.9\linewidth]{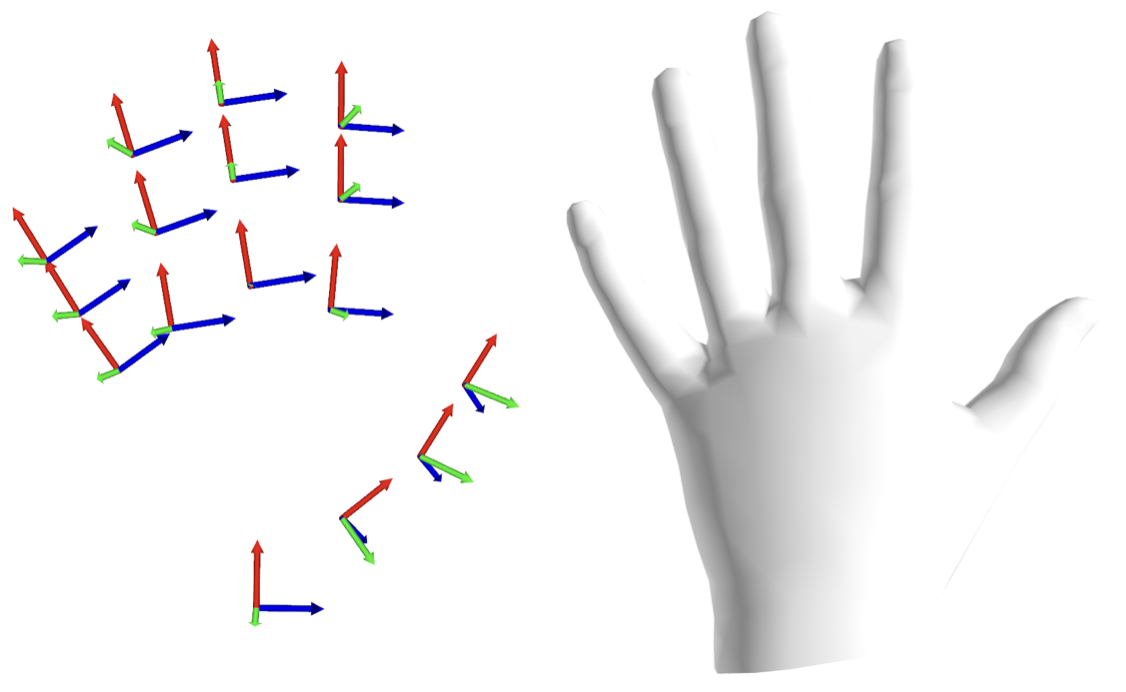}
    \caption{Local joint coordinate systems defined for the MANO model.}
    \label{fig:local_coordinates}
\end{figure}

To ensure biomechanical plausibility, we define local coordinate systems for the joints of the MANO model following \cite{spurr2020weakly}, as shown in \fig\ref{fig:local_coordinates}. We parameterize hand pose using anatomically informed degrees of freedom (DoF): the thumb’s metacarpophalangeal (MCP) and proximal interphalangeal (PIP) joints are assigned 3 DoF each, while its distal interphalangeal (DIP) joint is modeled with 1 DoF. For the other fingers, each MCP joint is also assigned 3 DoF, and both PIP and DIP joints are assigned 1 DoF. This parameterization yields a reduced pose vector $\bm{\phi} \in \mathbb{R}^{27}$, which is mapped in a differentiable manner to the full MANO pose space $\theta \in \mathbb{R}^{45}$. The global hand translation is represented as $\bm{t} \in \mathbb{R}^3$, and the global orientation $\bm{o}$ is modeled using a 6D rotation representation~\cite{zhou2019continuity}. 

Before capturing data for a new subject, we first estimate the hand shape parameters $\beta \in \mathbb{R}^{10}$ by fitting the MANO model to a 3D scan of the subject’s hand. This scan does not need to be high-precision and can be acquired using an affordable 3d scanner. In our experiments, we employed a low-cost \textit{Structure Sensor 1} depth camera mounted on an iPhone and operated with its official scanning application under default settings, which yields a coarse mesh (around 6k vertices) without any additional post-processing. The fitting is performed by minimizing the Chamfer Distance between the model and the scan, with optional supervision from direct finger length measurements. We optimize $\beta$ using gradient descent. Once estimated, $\beta$ is fixed for that subject and used in all subsequent motion capture sessions. 

Given the 3D marker data and fixed shape parameters $\beta$, we optimize the MANO model’s global translation $\bm{t}$, global orientation $\bm{o}$, and pose parameters $\bm{\phi}$ to fit the observed markers.

\begin{figure}[t]
    \centering
    \includegraphics[width=1.0\linewidth]{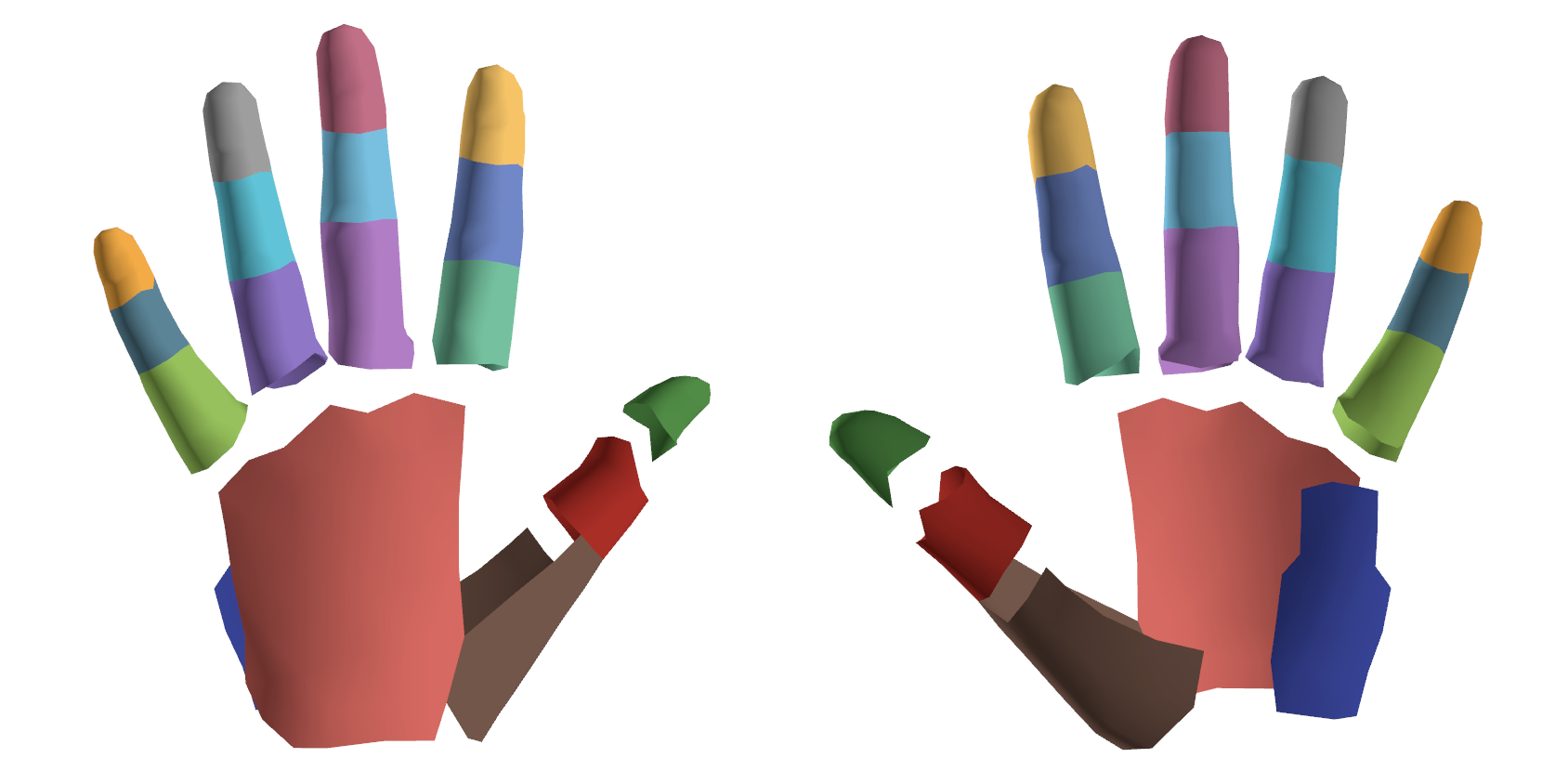}
    \caption{Manually defined submeshes of the MANO hand, with each color corresponding to a specific finger segment.}
    \label{fig:mano_segments}
\end{figure}

\subsubsection{Calibration}
\label{sssec:calibration}
Solving hand motion begins with a calibration process to associate the predefined markers with the MANO model, similar to prior non-rigid registration work\cite{white07capturing, halimi22pattern}. Formally, this association can be formulated as
\begin{equation}
\hat{\bm{x}}_i = M_i(\bm{t}, \bm{o}, \bm{\phi}, {b}) = b_{i}^{1} v_{i}^{1} + b_{i}^{2} v_{i}^{2} + b_{i}^{3} v_{i}^{3}
\label{eq:marker_barycentric}
\end{equation}
where $\hat{\bm{x}}_i$ represents the position of marker $i$, with the hat indicating that it is computed from the MANO model rather than measured from video. We assume that $\hat{\bm{x}}_i$ lies within the triangle formed by three mesh vertices $v_i^1$, $v_i^2$, and $v_i^3$, and its position is expressed using barycentric coordinates $b_i^{\{1,2,3\}}$ corresponding to each vertex. ${b}$ collectively represents all such associations and their corresponding barycentric parameters.

\begin{figure*}[t]
    \centering
    \includegraphics[width=0.95\linewidth]{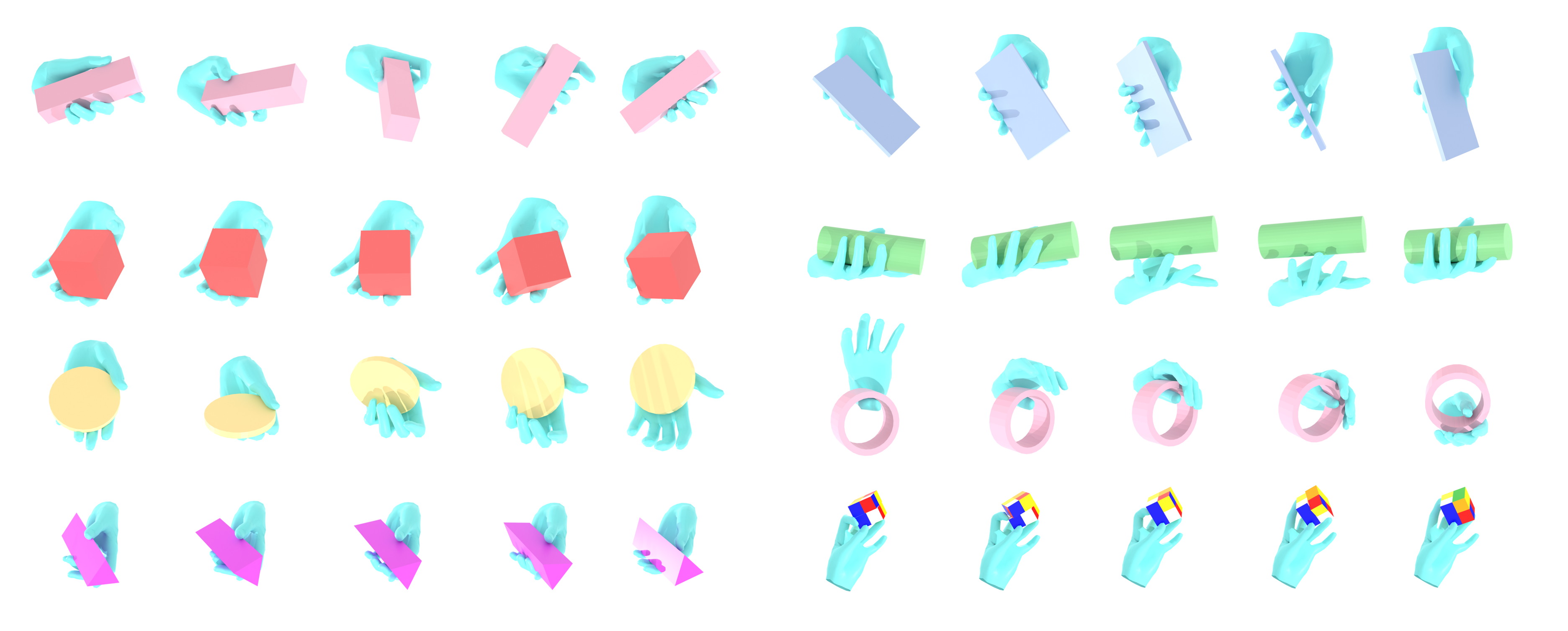}
    \caption{Hand-object interaction in the DexterHand dataset.}
    \label{fig:interaction}
\end{figure*}

To perform the calibration, we ask the subject to hold an open-hand pose similar to that shown in \fig\ref{fig:local_coordinates} and slowly rotate the hand several times so that all attached markers are seen by the cameras. This produces a calibration sequence on which we solve the association problem in Eq.~\eqref{eq:marker_barycentric} and Eq.~\eqref{eq:point_error_1} frame by frame.

In this process, we jointly optimize the global translation $\bm{t}$, global orientation $\bm{o}$, hand pose $\bm{\phi}$, and marker association parameters ${b}$ by minimizing the error between the predicted marker positions $M_i(\bm{t}, \bm{o}, \bm{\phi}, {b})$ and the captured 3D marker positions $\bm{x}_i$:
\begin{equation}
E_{\text{point}} = \sum_i \| M_i(\bm{t}, \bm{o}, \bm{\phi}, b) - \bm{x}_i \|_2^2
\label{eq:point_error_1}
\end{equation}
For the first frame, we initialize the parameters to zero and optimize \eqref{eq:point_error_1} using gradient descent for 1000 iterations. For each subsequent frame, we initialize from the previous solution. During optimization, we recompute the barycentric coordinates $b_i$ and their triangle indices using the nearest mesh triangle within the allowed submesh. This warm start speeds up convergence and yields temporally consistent marker-to-surface correspondences.

Because we know which body part each physical marker is attached to, we restrict the vertices $v_{i}^{1}, v_{i}^{2}, v_{i}^{3}$ for marker $i$ to lie within a predefined submesh $S_j$ of the MANO model corresponding to that finger segment (Fig.~\ref{fig:mano_segments}). These submeshes are predefined and serve as anatomical priors during optimization. Given a 3D marker point, we search only within $S_j$ and compute its closest point on the corresponding submesh by finding the triangle index and its barycentric coordinates $b_i$. This procedure can be interpreted as a segment-wise non-rigid ICP step with part-to-part correspondences between the marker cloud and the articulated MANO surface. Once calibration converges, the barycentric coordinates ${b}$ and their triangle indices are fixed and reused for motion reconstruction. We then optimize only the MANO pose parameters $\bm{\phi}$ and the global rigid motion $(\bm{t}, \bm{o})$.

\subsubsection{Solving Motion}
After calibration, the hand parameters ($\bm{t}, \bm{o}, \bm{\phi}$) for new motions are optimized on a per-frame basis, with each frame initialized from the solution of the previous one. The objective function in Equation~\eqref{eq:point_error_1} is now evaluated with $b$ fixed.

Additionally, a regularization term $E_{\text{reg}}$ is employed to penalize deviations from natural pose limits:
\begin{equation}
E_{\text{reg}} = \sum_{k\in N_{\text{DoF}}} \| \text{clip}(\bm{\phi}_k, \bm{\phi}_{k, \text{low}}, \bm{\phi}_{k, \text{high}}) - \bm{\phi}_k \|_2^2
\label{eq:reg_error}
\end{equation}
where the prior limits $\bm{\phi}_{\text{low}}$ and $\bm{\phi}_{\text{high}}$ are manually defined to cover the natural range of joint angles. 
The total objective function is then
\begin{equation}
E = E_{\text{point}} + \lambda_{\text{reg}} E_{\text{reg}}
\label{eq:total_objective}
\end{equation}
We optimize this objective function using Adam~\cite{Kingma2014AdamAM} optimizer with a learning rate of 0.002 for 400 epochs.

If a marker and all markers further along the same kinematic chain are occluded, we keep the corresponding joint DoFs at their previous-frame values. This improves temporal stability and reduces jitter. In our setting, occlusions during manipulation are usually brief, and the dense marker layout ensures that enough markers remain visible throughout the sequence. This makes the per-frame optimization reliable without requiring complex long-term occlusion handling.

\subsection{Object Pose Estimation}
\label{ssec:object_pose_estimation}

\begin{figure}[t]
    \centering
    \begin{subfigure}{0.49\linewidth}
        \centering
        \includegraphics[width=0.8\linewidth]{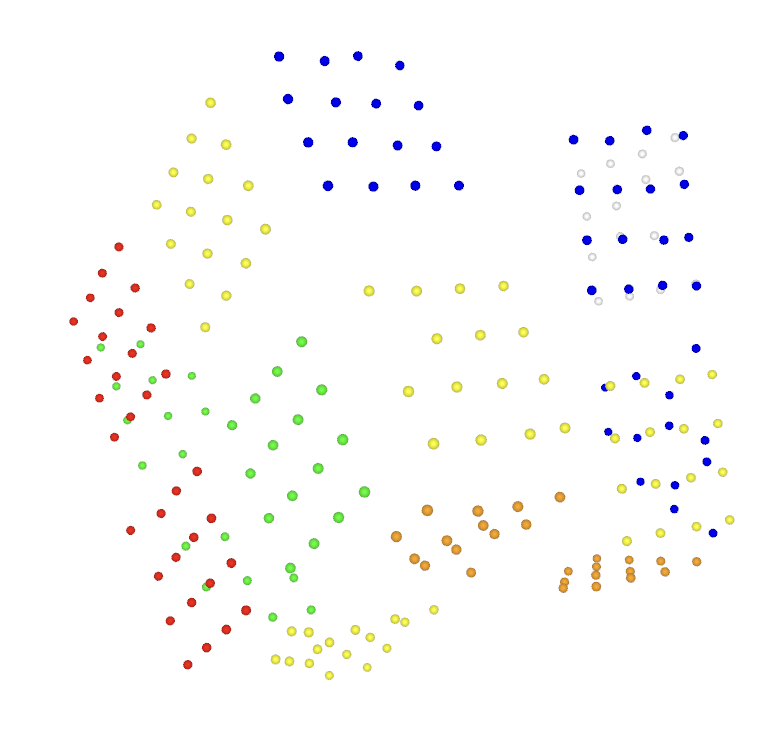}
        \caption{Marker points on a Rubik's cube.}
    \end{subfigure}
    \begin{subfigure}{0.49\linewidth}
        \centering
        \includegraphics[width=0.8\linewidth]{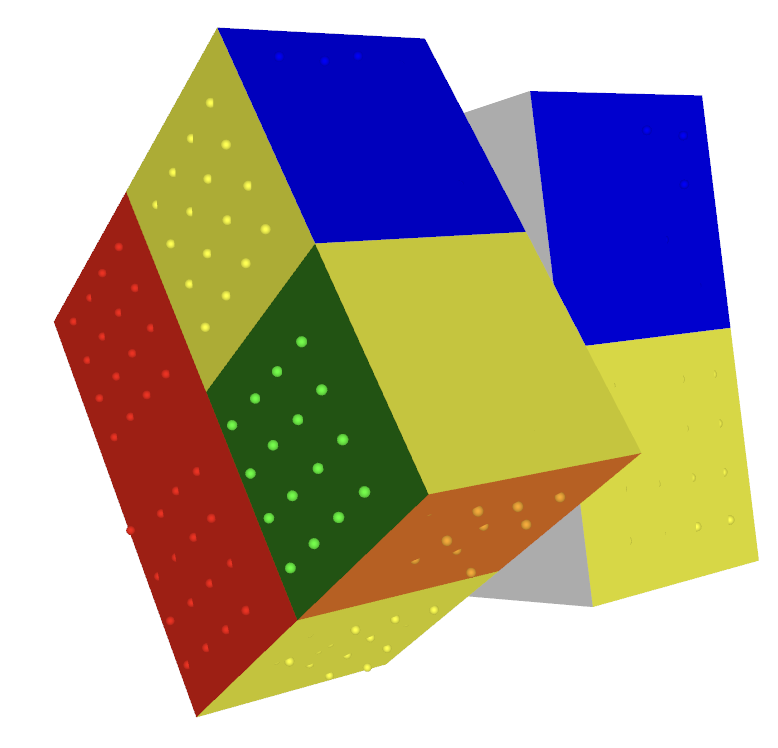}
        \caption{Rubik's cube pose estimation.}
    \end{subfigure}
    \caption{Object pose estimation.}
    \label{fig:obj_recon}
\end{figure}

\begin{table*}[t]
\centering
\caption{Overview of selected motion capture sequences in the DexterHand dataset. Shape dimensions and penetrations are in centimeters. Notations: $x$, $y$, $z$: length, width, and height of cuboid; $h$: height; $d$: diameter; $D$: outer diameter (ring);  $b$: side length of a triangular prism. }
\label{tab:dataset_stats}
\resizebox{0.7\linewidth}{!}{
\begin{tabular}{lccccc}
\toprule
Object Name & Object Shape (cm) & Total Time (s) & Penetration (cm) & MSNR $\uparrow$ & Jerk(m/s$^3$) $\downarrow$\\
\midrule
Cuboid      & $x=15.0, y=4.0, z=4.0$ & 473.65         & 0.46$\pm$0.33 & 9.17 & 0.99\\
Cuboid      & $x=16.0, y=0.8, z=7.5$ & 659.35         & 0.31$\pm$0.10 &9.29 & 0.67\\
Cuboid      & $x=6.0, y=6.0, z=6.0$ & 447.75 & 0.36$\pm$0.42 & 9.31 & 0.71 \\
Cylinder    & $d=5.5, h=16.0$ & 665.75 & 0.57$\pm$0.47 & 9.13 & 0.85\\
Disk & $d=10.0, h=1.0$ & 655.95 & 0.26$\pm$0.16 & 9.78 & 0.55 \\
Ring & $D=10.0, d=8.0, h=4.0$ & 700.70 & 0.31$\pm$0.17 & 9.06 & 0.93\\
Triangular Prism & $b=5.0, h=16.0$ & 585.70 & 0.40$\pm$0.36 & 9.24 & 0.88 \\
Rubik's Cube & $d=5.1$ & 747.75 & 0.36$\pm$0.22 & 9.50 & 0.46\\
\midrule
Total & $\varnothing$ & 4936.65 & 0.38$\pm$0.31 & 9.31 & 0.76\\
\bottomrule
\end{tabular}
}
\end{table*}

For sequences with object interaction, we estimate the object's 6-DoF pose (position and orientation) using known rigid 3D mesh models. We employ the Kabsch algorithm~\cite{kabsch1976solution} for all frames to align observed object markers $\bm{X}_{\text{obs}}$ with canonical marker positions $\bm{X}_{\text{canonical}}$. This algorithm determines the optimal rotation $\bm{R}_{\text{obj}}$ and translation $\bm{t}_{\text{obj}}$ by minimizing the following objective function:
\begin{equation}
L_{\text{obj}} = \| \bm{X}_{\text{obs}} - (\bm{R}_{\text{obj}} \bm{X}_{\text{canonical}} + \bm{t}_{\text{obj}}) \|_2^2
\end{equation}
where $\bm{R}_{\text{obj}}$ is the rotation matrix, and $\bm{t}_{\text{obj}}$ is the translation vector, computed directly for each frame to achieve precise alignment.

The reconstruction of the Rubik's Cube is relatively complex as shown in \fig\ref{fig:obj_recon}. Detailed implementation procedures are provided in Appendix~\ref{sec:rubik_cube_reconstruction}.

\section{DexterHand Dataset}
\label{sec:dataset}

To demonstrate the performance of our DexterCap system, we introduce \emph{DexterHand}, an open-source dataset that features a rich collection of dexterous hand-object interactions, with a particular focus on in-hand manipulation. 

We capture in-hand manipulations of seven basic object shapes, along with a Rubik’s Cube that features a compound articulated structure. Each capture session begins with a calibration sequence, as described in Section~\ref{sssec:calibration}, to establish reliable marker correspondences and ensure precise model initialization. Participants then perform a predefined set of actions, including basic free-hand movements (such as flexion, extension, abduction, and rotation) and a diverse array of in-hand manipulation skills.

Table~\ref{tab:dataset_stats} provides an overview of the current DexterCap dataset, summarizing the diversity of object types, geometric properties, and sequence durations. 
Prior to data collection, the subject provided informed consent for the acquisition and public release of the data. All released data are strictly anonymized and contain no personally identifiable information.

\section{Evaluation}
\label{sec:evaluation}

\subsection{System Setup}
\label{ssec:system_setup}
Our prototype system consists of 13 Hikvision MV-CS050-10GM industrial GigE PoE cameras, each capable of capturing grayscale video at 20\,FPS with a resolution of $2048 \times 2448$ pixels. The total hardware cost is under 6,000 USD. While using more or higher-end cameras can significantly improve performance, it also increases the overall cost. 

To train the models used for image processing and marker extraction, we manually label 2 to 3 frames from each video captured by each camera for every capture session. Our annotation focused on frames containing a high number of visible corners and blocks, resulting in an average of 241 corners and 132 blocks per frame. The annotation process is efficiently supported by a custom GUI-based tool developed for our system. The models are trained on this small dataset with standard image augmentations, including random color shifts, scaling, and warping. All models are optimized using the AdamW optimizer, with a typical learning rate of 0.001. We find that after training on manually labeled data from several capture sessions, totaling approximately 180 frames, the models generalize well and do not require retraining for new sessions recorded in the same environment. 

At runtime, the capture pipeline operates at 20\,FPS, while our unoptimized offline processing typically requires approximately 5\,s per frame for marker recognition and 5-12\,s per frame for hand-object reconstruction. Note that with the pretrained model described above, the offline processing pipeline is cascaded, fully automated, and can be executed in batch without additional human intervention. 

\subsection{Image Processing and Marker Recognition}
\label{ssec:recognition_performance}
To evaluate the performance of our marker detection and recognition pipeline, we use a manually annotated dataset and report several key metrics.

\textit{Corner detection}. Our CornerNet (heatmap-based) achieves 94.7\% precision, 81.6\% recall, and 87.7\% F1 score at the image level, where a corner is considered correctly detected if it lies within 5 pixels of the ground truth. In contrast, our implementation of the coordinate-regression model in \cite{chen2021capturing} reaches 98.3\% accuracy but only 46.3\% precision, yielding many false positives on patches without markers.

\textit{Edge recognition}. Our EdgeNet attains 99.02\% accuracy on the test set with 98.9\% precision, 99.1\% recall, and 99.0\% F1, indicating highly robust edge classification.

\textit{Block identification and ID recognition}. Our BlockNet achieves 98.39\% orientation accuracy, 97.95\% left-character accuracy, and 97.36\% right-character accuracy on the test set.  At the image level, our block recognition achieves a precision of 94.5\% with a recall of 41.3\%. The relatively lower recall is primarily attributable to the conservative strategies adopted in both corner detection and edge detection. While this may lead to true positives being undetected, it effectively reduces false positives and enhances robustness.

Notably, our edge detection model enables the edge-first assembly strategy to propose candidate quadrilaterals for block identification. Compared to the exhaustive quadrilateral validation strategy of \cite{chen2021capturing}, the edge-first assembly reduces the average number of candidates from 5550 quads per frame to 83 blocks (via ~707 edges), greatly improving efficiency and overall quality.

To show this, we reimplemented the method from \cite{chen2021capturing}. Although the model achieves block-level recognition accuracy exceeding 99\%, the extremely large number of quadrilateral candidates, most of which are negatives, substantially lowers its performance, resulting in only 55.0\% precision and 23.6\% recall.

\textit{Voting mechanism}. A voting mechanism based on inter-block adjacency further improves recognition accuracy. On average, 1.825\% labels are corrected by the voting mechanism.

\subsection{Hand-Object Reconstruction}
\label{ssec:reconstruction_accuracy}

To evaluate the accuracy of the reconstructed marker positions, we compute the reprojection errors and compare them to the reprojection errors obtained during the camera calibration process. Quantitatively, the triangulation of detected markers results in a reprojection error of {1.42 pixels}, in contrast to 0.4 pixels for camera calibration. Although our reprojection error is not as low as that of calibration, it is already quite satisfactory considering that calibration is performed with a perfect, rigid checkerboard under ideal conditions, whereas our markers are attached to small, deformable finger segments.

For reconstructing MANO parameters, we propose Marker Reconstruction Error (MRE), defined as the mean Euclidean distance between the predicted marker positions on the MANO model surface and the observed 3D marker positions. The resulting MRE is {0.77 $\pm$ 0.28 mm} (mean $\pm$ std) during calibration phases with high marker visibility, and {2.06 $\pm$ 1.09 mm} during dynamic manipulating phases with more complex motions and occlusions.

We estimate the 6-DoF pose of the object by minimizing position difference between the observed markers and target. The distance between each reconstructed object marker and its corresponding target position on the optimized rigid object model is {1.512 mm}. 

\subsection{Motion Quality and Plausibility}
\label{ssec:motion_quality}

We evaluate the motion quality and plausibility of our dataset with the following metrics. Heuristically, we apply a 5 Hz Butterworth low-pass filter during post-processing, described in detail in Appendix~\ref{sec:postprocessing}. The quantitative results are as follows.

\textit{Physical plausibility}. We evaluate the physical plausibility of our captured motion trajectories by measuring the penetration between the reconstructed hand model and object mesh. The average penetration is \textbf{3.8mm±3.1mm}, which is consistent with typical human hand deformation under realistic grasping forces, indicating good accuracy of our capture system.

\textit{Benchmark against other motion capture methods.}
We benchmark our dataset and pipeline against several public datasets: GRAB~\cite{taheri2020grab}, ARCTIC~\cite{fan2023arctic}, and HUMOTO~\cite{lu2025humoto}, which are captured using commercial marker-based systems or data gloves. For these datasets, we compute all metrics from the released trajectories using the same implementation as for our data. We also consider the vision-only methods HaMeR~\cite{pavlakos2024reconstructing} as an additional baseline. In particular, HaMeR is a single-view, vision-based method. We apply it to our captures by running it independently on one of our camera views. Since its performance can degrade on our grayscale videos, and because the markers attached to the fingers may alter the hand appearance relative to its training distribution, we further include the GigaHands~\cite{fu2025gigahands} dataset as an additional baseline. GigaHands is a multi-view extension of HaMeR, and we compute the same metrics on its released trajectories.

As shown in Table~\ref{tab:benchmark}, our method achieves competitive motion smoothness (low jerk) and reconstruction quality (high MSNR) compared with commercial systems, and substantially outperforms vision-only methods. Importantly, ours is the only system among these approaches that captures dexterous in-hand manipulation. Definitions of the quality metrics are provided in Appendix~\ref{sec:motion_quality_metrics}.

\begin{table}[t]
    \centering
    \caption{Benchmark comparison with vision-only and commercial systems. Bold values indicate best performance in each metric.}
    \label{tab:benchmark}
    \resizebox{\linewidth}{!}{
        \begin{tabular}{lcccc}
            \toprule
            Dataset            & MSNR $\uparrow$        & Jerk (m/s$^3$) $\downarrow$ & Diversity $\uparrow$ & Coherence $\uparrow$ \\
            \midrule
            Ours               & \textbf{9.31$\pm$0.22} & \textbf{0.76$\pm$0.18}      & \textbf{0.97}        & 0.68                 \\
            GRAB (Vicon)       & 7.29$\pm$1.51          & 3.68$\pm$1.74               & 0.91                 & 0.70                 \\
            ARCTIC (Vicon)     & 7.82$\pm$0.40          & 0.91$\pm$0.01               & 0.90                 & \textbf{0.81}        \\
            HUMOTO (Data Glove)    & 7.51$\pm$3.83          & 1.90$\pm$2.09               & 0.93                 & 0.63                 \\
            HaMeR (Vision)     & -0.05$\pm$0.07         & 23.76$\pm$0.78              & 0.90                 & \textbf{0.81}        \\
            GigaHands (Vision) & 3.50$\pm$1.23          & 2.62$\pm$1.42               & 0.91                 & 0.73                 \\
            \bottomrule
        \end{tabular}
    }
\end{table}

\section{Future Work and Conclusions}
\label{sec:conclusion}

In this work, we presented \emph{DexterCap}, a pipeline for capturing and reconstructing hand-object motion in dexterous manipulation. Our primary contributions include the development of a low-cost, high-precision motion capture system for hands and objects and the construction of a dedicated hand-object interaction dataset. 
The capture system, leveraging multiple cameras and robust marker-based tracking with outlier rejection and data processing, has demonstrated its capability to reconstruct fine-grained hand and object kinematics with high fidelity. This formed the foundation for our dataset, \emph{DexterHand}, which comprises various interaction tasks and provides a valuable resource for learning manipulation priors. 

As a vision-based method, our method remains vulnerable to severe occlusion. Because the pipeline relies on image-based marker detection, reconstruction quality degrades when a large fraction of markers are simultaneously invisible across all views. For example, in the extreme case where all fingers are inserted into the ring object shown in Figure~\ref{fig:interaction}, the hand becomes fully occluded, leading to artifacts such as finger-object penetration. Addressing this limitation likely requires occlusion-aware estimation, learning-based pose priors, or additional sensing modalities such as IMUs.

The current dataset, while valuable, can be expanded in scale and diversity. Capturing a wider range of subjects, objects (including deformable and articulated ones), and complex manipulation tasks, such as bimanual interactions and tool use, will significantly enrich the dataset. Augmenting the dataset with finer-grained semantic annotations, such as grasp types, functional intent, contact regions, and applied forces, would enable the training of more controllable and context-aware generative models.

Finally, integrating our data capture and generation pipeline with physics simulation environments offers exciting possibilities for validating motion plausibility and training robotic agents. Exploring sim-to-real transfer techniques to deploy policies on real robotic hands for complex manipulation tasks remains a challenging but rewarding goal. Continued research in these directions promises to advance our understanding of dexterous manipulation and enable new applications in robotics, and human-object interaction.
\appendix




\section{DANCE CARD FOR RUBIK'S CUBE}
\label{sec:dancecard}

\noindent The dance card serves as a structured protocol to ensure comprehensive motion coverage during data collection. This performer-facing guide systematically elicits diverse manipulation patterns while maintaining consistency across recording sessions. Note that the dance card functions solely as a data collection tool and does not constitute part of our optimization pipeline.

\begin{table}[h]
    \centering
    \caption{Dance Card for 2$\times$2$\times$2 Rubik's Cube Manipulation}
    \label{tab:2x2x2-dancecard}
    \resizebox{\linewidth}{!}{
    \begin{tabular}{cp{5cm}c}
        \hline
        \textbf{Step} & \textbf{Descriptions} & \textbf{Notation} \\
        \hline
        1 & Start in T-pose, hands relaxed, cube on table & $\varnothing$ \\
        2 & Reach out and grasp the 2$\times$2$\times$2 cube from left side with the left hand & $\varnothing$ \\
        3 & Lift the cube, no rotation & $\varnothing$ \\
        4 & Rotate the entire cube with wrist along the left-right (X) axis and rotate back 5 times & $(xx')5$ \\
        5 & Rotate the entire cube with wrist along the down-up (Y) axis and rotate back 5 times & $(yy')5$ \\
        6 & Rotate the entire cube with wrist along the back-front (Z) axis and rotate back 5 times & $(zz')5$ \\
        7 & Rotate the entire cube with manipulation along the left-right (X) axis clockwise and counter-clockwise 2 loops (8 times) & $(x4)2(x'4)2$ \\
        8 & Rotate the entire cube with manipulation along the down-up (Y) axis clockwise and counter-clockwise 2 loops (8 times) & $(y4)2(y'4)2$ \\
        9 & Rotate the entire cube with manipulation along the back-front (Z) axis clockwise and counter-clockwise 2 loops (8 times) & $(z4)2(z'4)2$ \\
        10 & Rotate the right face clockwise and counter-clockwise by 2 loops (8 times) & $(R4)2(R'4)2$ \\
        11 & Rotate the up face clockwise and counter-clockwise by 2 loops (8 times) & $(U4)2(U'4)2$ \\
        12 & Rotate the front face clockwise and counter-clockwise by 2 loops (8 times) & $(F4)2(F'4)2$ \\
        13 & Rotate the left face clockwise and counter-clockwise by 2 loops (8 times) & $(L4)2(L'4)2$ \\
        14 & Rotate the back face clockwise and counter-clockwise by 2 loops (8 times) & $(B4)2(B'4)2$ \\
        15 & Rotate the down face clockwise and counter-clockwise by 2 loops (8 times) & $(D4)2(D'4)2$ \\
        16 & Do an algorithm\footnotemark 4 times & $(RUR'U')6$ \\
        17 & Do random turns of the cube & $\varnothing$ \\
        \hline
    \end{tabular}
    }
\end{table}
\footnotetext{This algorithm is also known as the "Sexy Move".}

\section{RUBIK'S CUBE RECONSTRUCTION}
\label{sec:rubik_cube_reconstruction}

Our Rubik's cube reconstruction system employs a marker-based motion capture approach to track and reconstruct the state of a 2$\times$2$\times$2 Rubik's cube during manipulation. This section details the algorithmic framework and key technical components.

\subsection{System Overview}

The reconstruction pipeline processes motion capture data containing 3D positions of 384 markers (16 markers per facelet $\times$ 24 external facelets) to estimate the cube's global pose and internal rotation states. The algorithm operates on the principle that face rotations can be detected through coplanarity analysis of marker clusters, enabling decomposition of the complex 6-DOF object motion into manageable components.

\subsection{Marker Configuration and Reference Model}

Each of the 24 external facelets contains a 4$\times$4 grid of retroreflective markers with 5mm spacing. The reference cube model defines canonical marker positions in the solved state, with each marker assigned to one of 24 canonical facelet indices. The local marker positions within each facelet are generated using:
\begin{equation}
\bm{m}_{i,j} = \left[(i-1.5)s,\, (j-1.5)s,\, 0\right]
\end{equation}
where $s = 5$mm is the marker spacing, and $i,j \in \{0,1,2,3\}$ represent grid coordinates within the facelet.

\subsection{Coplanarity-Based Rotation Detection}

The core insight of our approach is that during face rotations, specific marker clusters exhibit characteristic coplanarity patterns. We define the coplanarity score for a set of 3D points $\mathcal{P}$ as the smallest singular value from Principal Component Analysis:
\begin{equation}
\text{Score}(\mathcal{P}) = \sigma_{\min}(\bm{P} - \bar{\bm{p}})
\end{equation}
where $\bm{P}$ is the $3 \times N$ matrix of centered point coordinates and $\sigma_{\min}$ denotes the smallest singular value.

For each of the six major faces (U, D, L, R, F, B), we monitor both the complete 4-facelet clusters and adjacent facelet pairs. A rotation is detected when two opposing face clusters remain coplanar (score $< 0.008$) while four orthogonal face clusters become non-coplanar (score $> 0.009$). This characteristic pattern emerges because the rotating face and its opposite maintain their planar structure, while the perpendicular faces undergo geometric distortion as their constituent facelets split between the moving and stationary portions of the cube.

\subsection{122-Block Decomposition and Registration}

Upon detecting a rotation, we decompose the cube into two 1$\times$2$\times$2 blocks based on the rotation axis. For example, during a U-face rotation, the lower block contains the down face and the bottom halves of the lateral faces (D, L$_{[2,3]}$, F$_{[2,3]}$, R$_{[2,3]}$, B$_{[2,3]}$), while the upper block comprises the up face and the top halves of the lateral faces (U, L$_{[0,1]}$, F$_{[0,1]}$, R$_{[0,1]}$, B$_{[0,1]}$). This decomposition isolates the moving portion from the stationary portion, enabling independent pose estimation for each component.

Each block is independently registered to its reference configuration using the Kabsch algorithm. For corresponding visible marker sets $\{\bm{p}_i\}$ and $\{\bm{q}_i\}$, the optimal rotation $\bm{R}$ and translation $\bm{t}$ minimize:
\begin{equation}
\min_{\bm{R},\bm{t}} \sum_{i} \|\bm{p}_i - (\bm{R}\bm{q}_i + \bm{t})\|^2
\end{equation}
The solution involves SVD of the cross-covariance matrix $\bm{H} = \sum_i (\bm{q}_i - \bar{\bm{q}})(\bm{p}_i - \bar{\bm{p}})^T$.

\subsection{Relative Rotation Estimation}

The internal rotation angle between the two 122-blocks is computed from their relative orientation:
\begin{equation}
\theta = \text{sign}(\bm{n}^T \cdot \boldsymbol{\omega}) \|\boldsymbol{\omega}\|
\end{equation}
where $\boldsymbol{\omega}$ is the rotation vector from $\bm{R}_{\text{rel}} = \bm{R}_2 \bm{R}_1^{-1}$, and $\bm{n}$ is the rotation axis vector transformed by $\bm{R}_1$.

\subsection{Implementation Details}

The reconstruction algorithm processes frames sequentially with a look-ahead window of 100 frames for rotation detection. The system employs coplanarity thresholds of $\tau_{\text{coplanar}} = 0.008$ and $\tau_{\text{non-coplanar}} = 0.009$ to distinguish between stationary and rotating configurations. To ensure discrete quarter-turn reconstructions, we apply an angle snapping threshold of $3^\circ$ that maps continuous rotations to discrete $90^\circ$ increments. The Kabsch registration requires a minimum of 3 visible markers per block for stable pose estimation, while marker occlusion is handled through placeholder coordinates $[-1000, -1000, -1000]$ that are excluded from geometric computations.

The algorithm accumulates incremental rotations until reaching $\pm 90^\circ$, then snaps to the nearest quarter-turn and updates the internal facelet index mapping accordingly. This approach ensures robust reconstruction even in the presence of noise and marker occlusions typical in real motion capture scenarios.

The complete reconstruction outputs frame-by-frame cube poses (translation, orientation) and face rotation parameters, enabling faithful replay of the manipulation sequence and downstream analysis of manipulation quality metrics.

\section{Model Architectures}
\label{sec:network_architecture}

\begin{table}[t]
    \centering
    \caption{Architecture of CornerNet.}
    \label{tab:unet_architecture}
    \setlength{\tabcolsep}{2.5pt}
    \renewcommand{\arraystretch}{0.9}
    \resizebox{0.98\columnwidth}{!}{%
        \begin{tabular}{llllll}
            \toprule
            Stage      & Operation             & Kernel     & Stride & Channels               & Notes           \\
            \midrule
            Enc-1      & (Conv+ReLU) $\times$2 & $3\times3$ & 1      & $1 \rightarrow 64$     & padding=1       \\
            Pool-1     & MaxPool               & $2\times2$ & 2      & $64 \rightarrow 64$    &                 \\
            Enc-2      & (Conv+ReLU) $\times$2 & $3\times3$ & 1      & $64 \rightarrow 128$   & padding=1       \\
            Pool-2     & MaxPool               & $2\times2$ & 2      & $128 \rightarrow 128$  &                 \\
            Enc-3      & (Conv+ReLU) $\times$2 & $3\times3$ & 1      & $128 \rightarrow 256$  & padding=1       \\
            Pool-3     & MaxPool               & $2\times2$ & 2      & $256 \rightarrow 256$  &                 \\
            Enc-4      & (Conv+ReLU) $\times$2 & $3\times3$ & 1      & $256 \rightarrow 512$  & padding=1       \\
            Pool-4     & MaxPool               & $2\times2$ & 2      & $512 \rightarrow 512$  &                 \\
            Bottleneck & (Conv+ReLU) $\times$2 & $3\times3$ & 1      & $512 \rightarrow 1024$ & padding=1       \\
            \midrule
            Dec-4      & Transposed Conv       & $2\times2$ & 2      & $1024 \rightarrow 512$ & upsampling      \\
            Dec-4      & (Conv+ReLU) $\times$2 & $3\times3$ & 1      & $1024 \rightarrow 512$ & skip from Enc-4 \\
            Dec-3      & Transposed Conv       & $2\times2$ & 2      & $512 \rightarrow 256$  & upsampling      \\
            Dec-3      & (Conv+ReLU) $\times$2 & $3\times3$ & 1      & $512 \rightarrow 256$  & skip from Enc-3 \\
            Dec-2      & Transposed Conv       & $2\times2$ & 2      & $256 \rightarrow 128$  & upsampling      \\
            Dec-2      & (Conv+ReLU) $\times$2 & $3\times3$ & 1      & $256 \rightarrow 128$  & skip from Enc-2 \\
            Dec-1      & Transposed Conv       & $2\times2$ & 2      & $128 \rightarrow 64$   & upsampling      \\
            Dec-1      & (Conv+ReLU) $\times$2 & $3\times3$ & 1      & $128 \rightarrow 64$   & skip from Enc-1 \\
            \midrule
            Output     & Conv                  & $1\times1$ & 1      & $64 \rightarrow 1$     &                 \\
            \bottomrule
        \end{tabular}
    }%
\end{table}

\noindent As summarized in Table~\ref{tab:unet_architecture}, CornerNet adopts a U-Net-based~\cite{ronneberger2015unet} encoder-decoder architecture with four resolution levels and symmetric skip connections. 

EdgeNet is built upon a ResNet-34~\cite{he2015resnet} backbone, adapted to single-channel inputs, and followed by a lightweight fully connected prediction head.

BlockNet is built upon a ResNet-34 backbone, adapted to single-channel inputs, and equipped with three parallel output heads that predict the label of two characters and the block orientation.

\section{Post Processing}
\label{sec:postprocessing}

The raw optimized motion parameters are post-processed to ensure temporal smoothness. A Butterworth low-pass filter is applied to the time series of global translation $\bm{t}$ (per dimension) and joint angles. For data captured at 20\,Hz, the cutoff frequency is typically set to 5\,Hz.

We choose a 5\,Hz cutoff based on empirical validation. Most voluntary dexterous hand motions occur below 5\,Hz, and in our experiments this cutoff preserves manipulation details while suppressing high-frequency noise. So 5\,Hz is a practical trade-off between noise reduction and motion fidelity for the hand-object interactions we capture.

\section{Motion Quality Metrics}
\label{sec:motion_quality_metrics}

\textit{MSNR}. A higher Motion Signal-to-Noise Ratio (MSNR) indicates smoother motion. Following HUMOTO~\cite{lu2025humoto}, we compute
\begin{equation}
\mathrm{MSNR}=10\log_{10}\left(\frac{\mathbb{E}[\hat{v}^2]}{\mathbb{E}[|v-\hat{v}|^2]}\right).
\end{equation}
where $v$ represents the local joint velocity and $\hat{v}$ is the smoothed velocity obtained through convolution operation with kernel size 3.

\textit{Jerk}. We assess motion smoothness via jerk, defined as the temporal derivative of joint acceleration; lower values indicate smoother kinematics.

\textit{Diversity} and \textit{Coherence}. To characterize structural variability, we cluster the hand-object poses of each trajectory into five clusters (k=5) and compute the mean diversity and coherence. Diversity reflects the breadth of motion patterns, whereas coherence quantifies intra-cluster compactness.

\bibliographystyle{eg-alpha-doi} 
\bibliography{main}       




\end{document}